\documentclass[runningheads]{llncs}
\usepackage[T1]{fontenc}
\usepackage{graphicx}
\usepackage[misc]{ifsym}
\usepackage{amsmath}
\usepackage{algorithm}
\usepackage{algpseudocode}
\usepackage{booktabs}
\usepackage{tabularx}
\usepackage{caption}
\usepackage{subcaption}

\newcommand{\corr}{(\Letter)}
\newcommand{\repeatthanks}{\textsuperscript{\thefootnote}}
\usepackage{mwe}

\begin{document}

\title{Covariate Ordered Systematic Sampling as an Improvement to Randomized Controlled Trials}

\titlerunning{COSS as an improvement to RCTs}

\author{
    Deddy Jobson\inst{1}\thanks{Both authors contributed equally.} \corr \and
    Li Yilin\inst{1}\repeatthanks \and
    Naoki Nishimura\inst{2,3}  \and
    Yang Jie\inst{1} \and
    Koya Ohashi\inst{1} \and
    Takeshi Matsumoto\inst{1} 
}

\authorrunning{Jobson et al.}

\institute{Mercari Inc., Tokyo, Japan
\email{\{deddy,y-li,j-yang,k-ohashi,takeshi.matsumoto\}@mercari.com}
\and
Recruit Co. Ltd., Tokyo, Japan
\and
University of Tsukuba, Tsukuba, Japan \email{nishimura@r.recruit.co.jp}
}

\maketitle              

\begin{abstract}
    The Randomized Controlled Trial (RCT) or A/B testing is considered the gold standard method for estimating causal effects. Fisher famously advocated randomly allocating experiment units into treatment and control groups to preclude systematic biases. We propose a variant of systematic sampling called Covariate Ordered Systematic Sampling (COSS). In COSS, we order experimental units using a pre-experiment covariate and allocate them alternately into treatment and control groups. Using theoretical proofs, experiments on simulated data, and hundreds of A/B tests conducted within 3 real-world marketing campaigns, we show how our method achieves better sensitivity gains than commonly used variance reduction techniques like CUPED while retaining the simplicity of RCTs.\footnote{Source Code: 
    https://anonymous.4open.science/r/COSS\_reasearch-4A3B}

\end{abstract}

\section{Introduction}

At the heart of all causal inference is the randomized controlled trial (RCT)\cite{stolberg_randomized_2004}. An RCT involves randomly allocating experiment units to two segments: the treatment and the control group. The change whose causal effect we want to measure is then effected on the treatment group. Finally, the difference in a metric of interest between the treatment and control groups is measured. It is taken to be the causal effect with some degree of certainty, depending on the number of units in the experiment. RCTs have found many applications in the industry\cite{kohavi_controlled_2009,siroker__2015} and have been deemed the gold standard by which decisions related to the design of products are evaluated.

Fisher\cite{fisher_statistical_1992} advocated using RCTs to measure causal effects because the prior methods to decide which units to treat based on experimenter knowledge led to biases. The key component of RCTs is the random allocation of units into treatment and control groups. Doing so ensures that no systematic biases arise that make the findings of the experiment moot. However, random allocation can cause imbalances in the number of units allocated between the two variants, contributing to the variance in the estimate of the treatment effect.

The more units in the experiment, the more accurate the estimate of the treatment effect. However, increasing the number of units incurs a greater cost and also potentially exposes more of them to unintentional bad experiences that naturally arise in the experiment stage since we do not know for sure in advance if the planned change is good. It is, therefore, advantageous to reduce the sample size as much as possible while maintaining accuracy.

Several methods exist to reduce the required sample size for a randomized controlled trial. The most popular is Control Using Pre-Experiment Data (CUPED)\cite{deng_improving_2013}. CUPED reduces the number of samples required by modifying the target metric to a similar one with a lower variance. Following CUPED, other methods like MLRATE\cite{guo_machine_2022} have been proposed in the same direction. 

One problem with the variance reduction approaches proposed above is that they involve either using statistical or machine learning models or changing the goal metric, which requires more engineering effort. This makes them harder to adopt in practice. Another shortcoming is that modifying the goal metric, as done by CUPED, is acceptable for A/B tests with a binary decision. In the case of marketing campaigns, however, we would like to instead estimate the treatment effect itself for Key Performance Indicator (KPI) reporting purposes. For that reason, modifying the metric can cause concern among stakeholders who are not as familiar with the workings of CUPED. 

We take a different approach to sample size reduction. Instead of modifying the target metric, we adopt a more deterministic sampling strategy. Our strategy allows one to reduce variance while retaining the simplicity of the RCT, where we just take the difference of the mean to be our estimate of the treatment effect. We call our method Covariate Ordered Systematic Sampling (COSS). 

Our contributions can be summarized as follows: 
\begin{itemize}
    \item We develop a novel method for variance reduction in A/B testing (which we call COSS). 
    \item We prove theoretically that it performs better than CUPED without using additional covariates or a more complicated algorithm.
    \item We perform simulation experiments to show where COSS achieves variance reduction when CUPED cannot. 
    \item We conduct several marketing campaigns using RCT and COSS to allocate users into treatment and control groups, finding COSS to greatly improve the decision power of our A/B tests and also lead to more accurate reporting. 
\end{itemize}


\section{Problem Statement}

We now formalize our problem statement. Given we have $2N$ experimental units, where unit $i$ has an associated outcome $Y_i$ and an informative covariate $X_i$, our goal is to estimate the average causal effect $\Delta$ of treatment on $Y$. In an RCT setting, users are randomly allocated to treatment ($T$) and control ($C$) groups, and the causal effect is estimated to be the difference of the means.

\begin{align}
    \Delta &= \frac{1}{2N}\left(\sum_{i\in T}Y_i - \sum_{i\in C}Y_i \right)
\end{align}




\section{Proposed Method}
\subsection{Our Method}

In COSS, we order covariates in descending order of $X$ and alternatively allocate samples to be in treatment and control groups (Algorithm \ref{alg:coss}). Without loss of generality, we assume that
\begin{align}
    X_i &\ge X_j \quad \forall i \le j \label{eq:sys_ineq}\\
    T &= \{0,2,4,\dots\} \\
    C &= \{1,3,5,\dots\} 
\end{align}

\begin{algorithm}
\caption{Covariate Ordered Standardized Sampling (COSS)}
\label{alg:coss}
\begin{algorithmic}[1]
\State Consider $2N$ elements with indices $i = 0, \dots, 2N-1$
\State Let $X$ be the pre-experiment covariate we use for COSS. 
\State Order elements in descending order of $X$ such that $X_i \ge X_j$ for all $i \le j$
\For{$i = 0, \dots, 2N-1$}
    \If{$i \bmod 2 = 0$}
        \State Assign element $i$ to Treatment group ($T$)
    \Else
        \State Assign element $i$ to Control group ($C$)
    \EndIf
\EndFor
\State \Return Experiment groups $T$ and $C$
\end{algorithmic}
\end{algorithm}

\subsection{Estimated Bias}
One limitation of our sampling methodology is that there is a systematic bias. We show theoretically that the bias induced by our sampling method is negligible and later show empirically that the bias does not adversely affect the type-1 error rate.

Without loss of generality, we let $Y$ and $X$ be unit-normalized. We can model their relationship in the following way: 
\begin{equation}
    Y = f(X) + \epsilon 
\end{equation}
where $f(\cdot)$ is any monotonically increasing or decreasing function, and $\epsilon$ is a random variable that captures the prediction error. We choose $f(\cdot)$ to minimize the variance of $\epsilon$. We do not need to know the exact formulation of $f(\cdot)$, but note that it results in the variance of $epsilon$ being less than or equal to that obtained from an affine $f(\cdot)$. While this model is sufficient to prove that COSS can achieve better performance than CUPED, smooth non-monotonic functions can be expressed as piecewise monotonic functions. So, we expect COSS to be effective in the presence of complex non-monotonic relationships, too. Let $r^2$ ($<1$) be the variance of $f(X)$. Our goal here is to prove that the bias induced by COSS is sufficiently small with a large enough sample size. 

\begin{proof}
Starting with the expression for $\Delta$:
\begin{align}
    \Delta &= \frac{1}{2N}\left(\sum_{i \in T} Y_i - \sum_{i \in C} Y_i \right) \nonumber \\
           &= \frac{1}{2N}\left(\sum_{i = 0}^{N - 1} (Y_{2i} - Y_{2i + 1}) \right) \nonumber \\
           &= \frac{1}{2N}\left( \sum_{i = 0}^{N - 1} (f(X_{2i}) - f(X_{2i + 1})) + \sum_{i = 0}^{N - 1} (\epsilon_{2i} - \epsilon_{2i + 1}) \right) \label{eq:delta_whole}
\end{align}
The bias is then the expected value of $\Delta$:
\begin{align}
    E[\Delta] &= \frac{1}{2N}\sum_{i = 0}^{N - 1} E[f(X_{2i}) - f(X_{2i + 1})] \nonumber \\
     &= \frac{1}{2N}\left( E[f(X_0)] - E[f(X_{2N-1})] - \sum_{i = 1}^{N - 1} E[f(X_{2i-1}) - f(X_{2i})]\right) \nonumber \\
  &\le \frac{1}{2N}(E[f(X_0)] - E[f(X_{2N-1})]); \quad (E[f(X_{2i-1}) - f(X_{2i})] \ge 0 \quad \forall i: \text{by Equation } \ref{eq:sys_ineq})\nonumber \\
\end{align}
Given that $X$ is standardized with zero mean and unit variance, the bias can be upper bounded by $\frac{E[\max_i f(X_i)]}{N}$. Different distributions of $f(X)$ yield different proportionalities for the bias:
\begin{itemize}
    \item Uniform: $\frac{1}{N}$
    \item Normal: $\frac{\sqrt{2\log{N}}}{N}$
    \item Shifted Poisson: $\frac{\log{N}}{N \log{\log{N}}}$
\end{itemize}
Given that the standard deviation of estimates of the mean decreases by a factor of $\frac{1}{\sqrt{N}}$, we conclude that the bias diminishes more rapidly than the variance, indicating its insignificance as $N$ increases.
\end{proof}

\subsection{Estimated Variance Reduction}

We now estimate the variance reduction achieved through COSS using Equation \ref{eq:delta_whole}.

\begin{align}
    Var(\Delta) &= Var\left( \frac{1}{2N}\left( \sum_{i = 0}^{N - 1} (f(X_{2i}) - f(X_{2i + 1})) + \sum_{i = 0}^{N - 1} (\epsilon_{2i} - \epsilon_{2i + 1}) \right) \right) \\
    &= \frac{1}{4N^2}\left( Var\left( \sum_{i = 0}^{N - 1} (f(X_{2i}) - f(X_{2i + 1}))\right) + Var\left(\sum_{i = 0}^{N - 1} (\epsilon_{2i} - \epsilon_{2i + 1}) \right) \right) \\
    &= \frac{1}{4N^2} Var\left( \sum_{i = 0}^{N - 1} (f(X_{2i}) - f(X_{2i + 1}))\right) + \frac{1}{2N}Var(Y)(1-r^2) \\
    &\le \frac{1}{4N^2} (Var(f(X_0)) + Var(f(X_{2N-1}))) + \frac{1}{2N}Var(Y)(1-r^2) 
\end{align}

If $X\sim Normal(0,1)$, then $Var(X_0)$ is of $O(\frac{1}{\sqrt{n}})$\cite{granville_beautiful_2019}. Therefore, the first term is of $O\left(\frac{1}{\sqrt{n^3}}\right)$ and so decays faster than the second term. 


We have two terms, one with a denominator of $N$ (similar to that of CUPED) and another with a denominator of $N^2$. As we can see, the variance reduction achieved is similar to that of CUPED, save for an additional rapidly decreasing constant value. However, if $X\perp Y$, then following the logic laid out by Athey et al. for stratified sampling\cite{athey_econometrics_2016}, the performance of COSS becomes equivalent to that of an RCT. 

\section{Related Work}

We now go through some of the existing work towards reducing the samples required to conduct online experiments. Reducing the number of samples reduces the cost of running the experiment and also reduces the number of bad treatments inadvertently given in case a questionable experiment is run. Statisticians have worked on reducing the error of estimations from RCTs for a long time and have come up with several methods. Regression adjustment\cite{rubin_use_1973,lin_agnostic_2013} is the most standard method in such a case. The idea is to model the target variable as an affine transformation of the covariate. 
\begin{equation}
    Y = \beta_0 + \beta_1 X + \epsilon
\end{equation}
In the above formulation, $\epsilon \sim N(0,\sigma^2)$ and $\epsilon \perp X$. The treatment effect is estimated through $\beta_1$, and we can achieve a variance reduction proportional to the square of the correlation between $X$ and $Y$. 

The linear regression-adjusted method can be extended to non-linear methods too, as done by Negi et al.\cite{negi_revisiting_2021}, or with multiple covariates, as done by Freedman\cite{freedman_regression_2008}. However, concerns have been raised by authors, including Rosenblum et al.\cite{rosenblum_using_2009}, that the validity of regression adjustment methods hinges on the relationship between covariate and target being as specified by the model. To circumvent such violations of assumptions, semi-parametric methods\cite{tsiatis_covariate_2008} have also been developed. 

Deng et al.\cite{deng_improving_2013} propose Covariate adjustment Using Pre Experiment Data (CUPED), a method similar to regression adjustment. Since their method does not assume any particular relationship between the covariate and target, they avoid some of the pitfalls of regression adjustment methods. CUPED works by creating a new target variable. 
\begin{equation}
    Y \leftarrow Y - \theta (X - E[X])
\end{equation}
Since the additive adjusted term has an expected value of zero, the adjusted target variable has the same expectation as the original target variable for any value of $\theta$. We set $\theta$ to minimize the variance of the adjusted target variable. Deng et al.\cite{deng_improving_2013} found that the achieved variance reduction is proportional to the square of the correlation coefficient between the covariate and the target variable. CUPED has found widespread adoption in the industry\cite{xie_improving_2016} and has spawned related methods to improve the sensitivity of A/B tests by related methods like in-experiment data\cite{deng_variance_2023}, variance-weighting\cite{liou_variance-weighted_2020}, one-sided-triggering\cite{deng_zero_2023}, etc. Another extension of CUPED is to use multiple variables, as done by the authors of MLRATE\cite{guo_machine_2022}. 

Another method similar in spirit to CUPED is stratified sampling. In stratified sampling, the data are first divided into strata using a categorical covariate like gender, city of residence, etc. Within each stratum, simple randomized controlled trials are conducted. During evaluation, the data are first aggregated within each stratum before being aggregated between strata. This reduces the variance if the strata are chosen such that the within-stratum variances are smaller than the between-strata variances. 

The abovementioned methods benefit from being applied post-experiment; they are safe since the original experiment setting is not tampered with. The next class of methods we explore involves changing the sampling of experiment units to improve the sensitivity of RCTs. This will result in there being a single answer to what the treatment effect is while guaranteeing improved sensitivity. 

\begin{figure}
    \centering
    \includegraphics[width=1\linewidth]{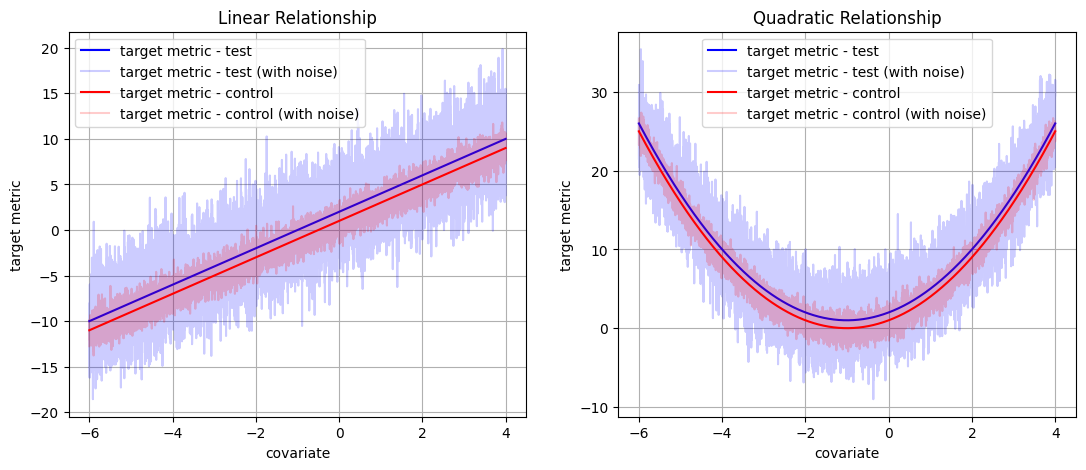}
    \caption{the distribution of simulation experiments.}
    \label{fig:simulation_experiment_distributions}
\end{figure}

Systematic sampling\cite{iachan_systematic_1982} is the method most similar to our method wherein users are, just like in COSS, alternatively chosen to be in the treatment and control group. It is a statistical method from survey methodology used for convenience when the population is relatively homogenous, for example, when surveying households or customers entering a supermarket\cite{beynon_gender_2010}. A crucial difference between systematic sampling and COSS is that in systematic sampling, we assume the underlying covariate (geographical location, for example) is independent of the outcome variable to get an unbiased estimate of the changes in the outcome variable. In COSS, however, we prefer to choose the covariate by which we order the experiment units before sampling so that it has a high correlation with the target variable. Therefore, our method achieves variance reduction in exchange for a provably small bias. 

Adaptive sampling\cite{pocock_sequential_1975} has been touted as a method to help balance covariates in experiments in medicine. They have especially found use when working with low sample sizes. A special case of adaptive sampling, especially similar to COSS, is the matched pair design\cite{imai_variance_2008,bai_optimality_2022}. However, all adaptive sampling methods proposed have worked with data processed one by one in random order; our method is different in that we sort all users deterministically before allocation. Because of this difference, we get simpler derivations on the bound of the expected variance reduction and the bias. 

A common hurdle to applying any of the above methods is its complexity. COSS offers an alternative that proves just as powerful (or more) to improve the sensitivity of RCTs while being trivial and intuitive to implement.

\begin{table}
\centering
    \begin{tabular}{p{0.3\linewidth}p{0.25\linewidth}p{0.25\linewidth}}
    \toprule
    Strategy & mean &  standard error \\
    \midrule
    RCT(original) & 0.998 & 0.874 \\
    RCT(with CUPED) & 1.000 & 0.313 \\
    COSS & 0.899 & 0.318 \\
    \bottomrule\\
    \end{tabular}
\caption{Linear Relationship - mean and standard error of treatment effect. COSS and CUPED exhibit comparable efficacy, both reducing the standard error by 65\%.}
\label{tab:linear_relathionship_stats}
\end{table}

\section{Experiments}

We conducted several experiments. Offline experiments were primarily conducted using simulated data, and online experiments were conducted in marketing campaigns at a large C2C marketplace with millions of active users. 
Our primary goals are as follows:
\begin{itemize}
    \item To measure the extent of variance reduction achievable through COSS. 
    \item To verify empirically that paired t-tests should be conducted in place of independent t-tests when performing hypothesis testing with COSS. 
    \item To check the extent to which COSS improves the power of the RCT without compromising on the type 1 error. 
\end{itemize}

\begin{figure}
    \centering
    \includegraphics[width=0.95\linewidth]{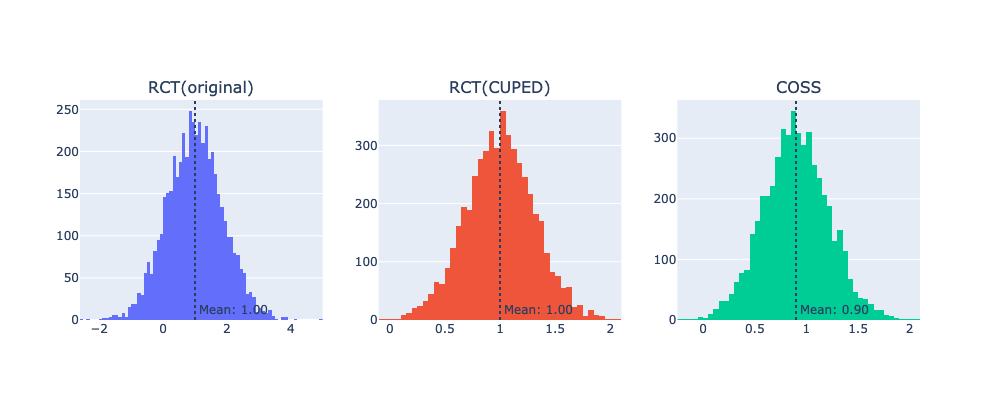}
    \caption{Linear Relationship: the distribution of treatment effect. Both CUPED (middle) and COSS(right) reduced variances significantly compared to RCT(left)}
    \label{fig:distribution_sample_mean_linear_relationship}
\end{figure}

\begin{figure}
    \centering
    \includegraphics[width=0.95\linewidth]{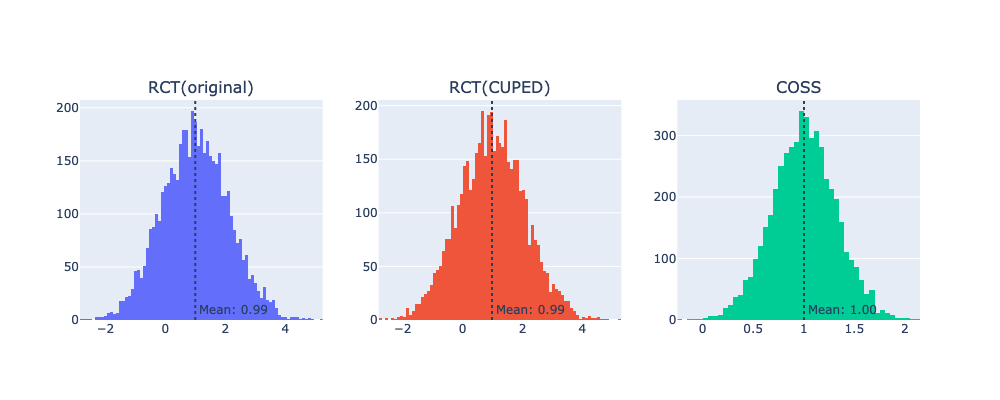}
    \caption{Quadratic Relationship: the distribution of treatment effect. COSS(right) reduced variances significantly compared to RCT(left) and CUPED (middle).}
    \label{fig:distribution_sample_mean_quadratic_relationship}
\end{figure}

\subsection{Simulation Experiments}
To assess the accuracy of COSS's ability to estimate the ground truth and its performance across various relationships (both linear and non-linear) between covariate and target metrics, we devised simulation experiments featuring diverse distributions. These simulations were then compared to the outcomes of RCT and CUPED.
\begin{itemize}
  \item simulation 1: linear relationship
\end{itemize}
\begin{equation}
y_{0} = bx + c + \mathcal{N}(0, \epsilon_0)
\end{equation}
\begin{equation}
y_{1} = bx + c + \mathcal{N}(\mu, \epsilon_1)
\end{equation}
\begin{itemize}
  \item simulation 2: quadratic relationship
\end{itemize}
\begin{equation}
y_{0} = ax^{2} + bx + c + \mathcal{N}(0, \epsilon_0)
\end{equation}
\begin{equation}
y_{1} = ax^{2} + bx + c + \mathcal{N}(\mu, \epsilon_1)
\end{equation}

Given that $x$ is covariate, $y_{0}$ is the target metric without treatment effect, and $y_{1}$ is the target metric with treatment effect. Set the treatment effect as a Normal distribution of the lift effect $\mu$. 

In our simulation experiments, we assigned the values $a = 1$, $b = 2$, $c = 1$, $\mu = 1$, $\epsilon_0 = 1$, $\epsilon_1 = 3$. We set a population dataset consisting of 10,000 data points. The visualization will be like fig \ref{fig:simulation_experiment_distributions}.


\subsubsection{Distribution of the average causal effect $\Delta$}
We randomly sampled 200 data points from the population 5,000 times. We applied COSS to allocate the variants for each sample and calculated the average effect $\Delta$. We did the same with RCT and CUPED settings. The distribution of treatment effect is plotted in Graphs \ref{fig:distribution_sample_mean_linear_relationship} and \ref{fig:distribution_sample_mean_quadratic_relationship}, and the results corresponding to linear and quadratic relationships are presented separately in Tables \ref{tab:linear_relathionship_stats} and \ref{tab:quadratic_relathionship_stats}, respectively. From the standpoint of reducing variance, COSS demonstrates equivalent capability to CUPED when examining linear relationships between variables. However, in scenarios where a quadratic relationship exists and the correlation between variables is near zero, CUPED is ineffective. In contrast, COSS not only matches the ground truth more closely but also significantly reduces the variance, showcasing its robustness in managing complex relationships. Upon evaluating the average of the estimated treatment effects in the linear case, we note that COSS introduces a small bias. This outcome aligns with what we proved in the earlier section; the observed bias is much smaller than the standard error.

\begin{table}
\centering
    \begin{tabular}{p{0.3\linewidth}p{0.25\linewidth}p{0.25\linewidth}}
    \toprule
    Strategy & mean &  standard error \\
    \midrule
    RCT(original) & 0.987 & 1.119 \\
    RCT(with CUPED) & 0.987 & 1.115 \\
    COSS & 1.002 & 0.318 \\
    \bottomrule\\
    \end{tabular}
\caption{Quadratic Relationship - mean and standard error of treatment effect. the standard error saw a reduction of 70\% in COSS, whereas CUPED did not demonstrate a substantial difference when compared to the results of RCT}
\label{tab:quadratic_relathionship_stats}
\end{table}



\subsubsection{Which t-test should we apply for COSS?}
As COSS organizes all samples based on covariate order and assigns them alternately to treatment and control groups, its sampling methodology resembles finding a twin for each sample and comparing the target metric—an approach akin to paired sample t-tests. We conducted paired sample t-tests and paired bootstrap sampling for COSS while employing independent sample t-tests and bootstrap sampling for RCT. Even though studies have shown that paired t-tests can result in inconsistencies\cite{ghadhban_robust_2021}, we observed consistent outcomes across all analyses. 

\begin{table}
\centering
    \begin{tabular}{p{0.35\linewidth}cc}
    \toprule
    Strategy & p-value of t-test & p-value of bootstrap sampling \\
    \midrule
    RCT & 0.2729 & 0.272 \\
    COSS & 0.0142 & 0.014 \\
    \bottomrule\\
    \end{tabular}
\caption{Estimations of the p-value of COSS and RCT using t-test and bootstrap sampling methods. The t-test result is in line with the bootstrap sampling result.}
\label{tab:similation_experiment_p-value_of_t_test_and_bs}
\end{table}

\subsection{Online Experiments}
Our business is a CtoC marketplace where customers can buy or sell items to each other. We conduct marketing campaigns on a regular basis to incentivize segments of customers to buy or sell more items. We use RCTs to evaluate the effectiveness of our marketing campaigns. To evaluate the effectiveness of COSS, we further segment users into those who are partitioned between treatment and control groups by COSS and those by RCTs.

We conducted an online marketing campaign targeting 12 million sellers in October 2023. The campaign was conducted over four segments of users based on historic levels of activity and with four different types of coupons: 5\% discount coupons, 10\% discount coupons, 30\% point back coupons, etc.  For our experiment, we also further allocated users equally to the two sampling methods: RCT and COSS. The covariate we used in COSS was the number of days since an item was put up for sale. 

We consider the following user-centric KPIs:
\begin{itemize}
    \item \textbf{lister:} whether or not the user put at least one item for sale during the campaign period. 
    \item \textbf{seller:} whether or not the user got at least one listed item sold during the campaign period. 
    \item \textbf{listings:} number of items put for sale during the campaign period. 
    \item \textbf{sold\_items:} number of items sold during the campaign period. 
    \item \textbf{gmv:} gross merchandise value (total amount) of items sold during the campaign period. 
\end{itemize}

This gave us ($4\: segments\times4\: coupon\times5\: metrics=$) 80 A/B tests, each further divided into two sampling types: RCT and COSS. Applying CUPED to those conducted using RCT data, we get another 80 results, which gives us a total of ($80\times3\: sampling \:methods=$) 240 A/B tests.  

\begin{table}
\centering
    \begin{tabular}{p{0.35\linewidth}cc}
    \toprule
    Sampling Strategy & \# Experiments & $|p|<0.05$ \\
    \midrule
    RCT & 80 & 13 \\
    CUPED & 80 & 14 \\
    COSS & 80 & 23 \\
    \bottomrule\\
    \end{tabular}
\caption{Summary of the results of 240 online experiments: our method COSS is the most sensitive approach.}
\label{tab:p_value_summary_ab_test1}
\end{table}

\begin{table}
\centering
    \begin{tabular}{p{0.35\linewidth}cc}
    \toprule
    Sampling Strategy & \# Experiments & $|p|<0.05$ \\
    \midrule
    RCT & 96 & 2 \\
    CUPED & 96 & 1 \\
    COSS & 96 & 2 \\
    \bottomrule\\
    \end{tabular}
\caption{Summary of the results of 288 AA tests: we see no substantial difference between the methods.}
\label{tab:p_value_summary_aa_test1}
\end{table}

To demonstrate the improvement in sensitivity of the A/B tests, we looked at the number of experiments with statistically significant lifts (Table \ref{tab:p_value_summary_ab_test1}). We see that more results from COSS turned out to be statistically significant. To make sure that the type 1 error rate of COSS is within limits, we also performed an AA test with the same set of users. We use common buyer-related and seller-related metrics to check if there is any substantial difference between them. Instead of looking at metrics measured during the campaign period, we measure them before the campaign period. Looking at Table \ref{tab:p_value_summary_aa_test1}, we see that COSS does not yield considerably more false positives than RCTs. While p-values are not directly used in KPI reporting, they can help answer certain marketing-product-related questions like "Which coupon type is most effective?" to improve further the effectiveness of future iterations of our marketing campaigns.

\begin{table}
\centering
    \begin{tabular}{p{0.35\linewidth}cc}
    \toprule
    Sampling Strategy & \# Experiments & $|p|<0.05$ \\
    \midrule
    RCT & 20 & 12 \\
    CUPED & 20 & 13 \\
    COSS & 20 & 14 \\
    \bottomrule\\
    \end{tabular}
\caption{Summary of the results of 60 online experiments: our method COSS is marginally the most sensitive approach.}
\label{tab:p_value_summary_aa_test3}
\end{table}


\begin{table}
\centering
    \begin{tabular}{p{0.35\linewidth}cc}
    \toprule
    Sampling Strategy & \# Experiments & $|p|<0.05$ \\
    \midrule
    RCT & 30 & 2 \\
    CUPED & 30 & 3 \\
    COSS & 30 & 1 \\
    \bottomrule\\
    \end{tabular}
\caption{Summary of the results of 90 AA tests: our method COSS is marginally the most sensitive approach.}
\label{tab:p_value_summary_ab_test3}
\end{table}

For our second online experiment, we conducted a similar marketing campaign in December 2023. This time, we targeted a different 12 million users. This time, we have ($2\: segments\times1\: coupon\times5\: metrics=$) 10 A/B tests for each of the three A/B testing methods: RCT, CUPED, and COSS. We used the past number of items put for sale as our covariate in CUPED and COSS. Unfortunately, due to interference from other end-of-year marketing campaigns, we could not achieve as many significant effects as before. We only found one statistically significant effect, which was obtained in one of our COSS experiments. AA tests revealed no false positives. 

For our third experiment, we performed a category expansion campaign for 10 days in February 2024. We aimed to motivate 16 million users to buy items in new categories and get introduced to new experiences. 
We tracked the following KPIs: 
\begin{itemize}
    \item \textbf{buyer:} whether or not the user bought at least one item during the campaign period. 
    \item \textbf{transactions:} the number of items bought during the campaign period. 
    \item \textbf{gmv:} gross merchandise value (total amount) of items bought during the campaign period. 
    \item \textbf{LTV:} the lifetime value of all actions (buy, sell, etc.) the user conducts in the long term. 
\end{itemize}

In total, we have ($5\: segments\times1\: coupon\times4\: metrics=$) 20 A/B tests for each of the three A/B testing methods: RCT, CUPED, and COSS. We use the past LTV as our covariate for CUPED and COSS. From Table \ref{tab:p_value_summary_ab_test3}, while we find CUPED and COSS to perform equally well with metrics, we find COSS is especially better when the target metric is the GMV. This agrees with our intuition because the GMV is a complex metric likely to have a non-linear relationship with the covariate, which COSS (but not CUPED) can handle. As before, we confirmed the type-1 error rates through AA testing (Table \ref{tab:p_value_summary_aa_test3}).

Lastly, in Table \ref{tab:ab_test1_variances}, we estimate the variance of the treatment effect using the bootstrap method\cite{efron_bootstrap_1979}. We use data from our first online experiment for this purpose. We normalize the values by constant factors to hide business metrics. COSS achieves more variance reduction than CUPED in most of our A/B tests. This translates to more stable reporting of KPI results, which is very helpful in the context of our marketing campaigns.

\begin{table}
\centering
\begin{tabularx}{0.6\linewidth}{@{}llrrr@{}}
\toprule
segment & sampling strategy & GMV & listings & sales \\ \midrule
1 & CUPED & 5.16 & 5.61 & 4.90 \\
& RCT & 4.94 & 6.46 & 5.66 \\
& COSS & \textbf{3.38} & \textbf{5.11} & \textbf{4.89} \\
\midrule
2 & CUPED & 18.09 & 21.56 & 24.19 \\
& RCT & 18.35 & 19.22 & 24.43 \\
& COSS & \textbf{17.07} & \textbf{17.88} & \textbf{22.97} \\
\midrule
3 & CUPED & 13.44 & 14.77 & 13.57 \\
& RCT & 12.90 & 17.51 & 11.30 \\
& COSS & \textbf{5.14} & \textbf{12.06} & \textbf{8.93} \\
\midrule
4 & CUPED & \textbf{16.14} & 4.52 & 20.29 \\
& RCT & 17.41 & 4.09 & 22.41 \\
& COSS & 16.41 & \textbf{1.93} & \textbf{19.54} \\ \bottomrule\\
\end{tabularx}
\caption{Normalized variance of the treatment effect (estimated using 200 bootstrap iterations) across different user segments and A/B testing methods from our first online experiment for the high-variance metrics. COSS performed best in almost all cases. Note that CUPED sometimes did even worse than RCT due to the high variances in estimating the target metrics. }
\label{tab:ab_test1_variances}
\end{table}

\section{Applying COSS in Practice}
In this section, we stray from the theoretical aspects of COSS and focus on the hurdles we faced implementing it in practice.

\subsubsection{Covariate selection}
To achieve the best improvement of power, it would be prudent to choose the covariate for ordering such that it has the strongest relationship with the target variable. The authors of CUPED\cite{deng_improving_2013} give a good starting point. They suggest using the same metric as the target metric but calculated using historical data for each user. We did the same in our online experiments and found the performance gains to be sufficiently good for COSS. 

\subsubsection{Convincing our business stakeholders to adopt COSS}
Variance reduction methods like CUPED have been applied in many companies to improve the sensitivity of A/B tests. They have primarily been used when binary decisions need to be made: for example, to deploy or to not deploy a feature. In such a scenario, the estimated treatment effect itself is not as important as the statistical significance of the result. In the context of marketing, the goal is typically to meet certain key performance indicator (KPI) targets. Business stakeholders are interested in the value of the lift in the KPIs generated by marketing campaigns. 

When we first proposed the adoption of CUPED to improve performance, this raised two concerns. The first was that CUPED involved the modification of the target variable. This led to concerns that we may not have been measuring the KPIs that they desired. The second had to do with the fact that CUPED is applied retroactively on a campaign. Since CUPED, while promising variance reduction, does not promise a smaller p-value (the estimated treatment effect may be smaller), there may be cases where simple RCTs give statistically significant results, which disappear with the adoption of CUPED. 

COSS helped us circumvent both concerns in one go. Firstly, COSS does not require the target variable to be modified; it only requires a different sampling method for the experiment units. It helps that the proposed sampling method is more intuitive, making it easier for stakeholders to accept. Secondly, since COSS is applied before the experiment occurs, we do not end up in the same situation as with CUPED, where we have two different results. While this is a trivial difference theoretically, it greatly helps reduce ambiguity and facilitates the practical adoption of COSS for our marketing experiments.  

\section{Limitations}
While COSS has several benefits over existing variance reduction techniques, 
there are some drawbacks to the current implementation of COSS. 
\subsubsection{Unequally-sized Variants}
Thus far, we have only explored how much variance reduction we can achieve by two variants of equal size. 
Finding a way to efficiently estimate the variance of the treatment effect with multiple unequally-sized variants is the next logical step for future research. 
\subsubsection{Multiple pre-experiment covariates}
Thus far, we have discussed how COSS can achieve variance reduction when we have one pre-experiment covariate. One potential way to extend COSS to multiple covariates is to train a machine-learning model to predict the target variable using all the covariates. The prediction variable from the model can be used as a proxy variable to apply in COSS. This is similar to the approach briefly mentioned by Guo et al.\cite{guo_machine_2022}. 
\subsubsection{Multiple outcomes of interest}
As noted in our various online marketing experiments, we may have multiple business KPIs we would like to track during the experiment. While post hoc methods can use different covariates for different target variables, COSS can only use one covariate before allocation. This is a fundamental limitation for which we believe there is no solution. However, COSS can be applied along with other methods like CUPED that do not have such hurdles to reduce the variance more than either method can on its own. 

\section{Conclusion}
In this paper, we propose a novel approach to improving the sensitivity of randomized controlled trials. We deployed our method in multiple online experiments and found that it yields more accurate estimations than standard industrial methods like CUPED. Given that our approach is different from existing methods that improve the sensitivity of RCTs by modifying the target variable, we can combine COSS with methods like MLRATE to achieve even more variance reduction than was thought possible. Furthermore, since we have only started to explore improving RCTs in this direction, we expect that a lot of improvements made to CUPED can also be applied to COSS, thus opening up a lot of possibilities for future research into improving the sensitivity of RCTs. 

%
%
%

\bibliographystyle{splncs04}
\bibliography{camera_ready}
%




\end{document}